\documentstyle[aas2pp4,graphics]{article}

\lefthead{Andersson, Kokkotas and Stergioulas}

\righthead{The r-mode instability and accreting compact stars}

\begin{document}

\title{On the relevance of the r-mode instability for accreting
neutron stars and white dwarfs}

\author{Nils Andersson}

\affil{Department of Mathematics, University of
Southampton, Southampton SO17 1BJ, UK\\
Institut f{\"u}r Astronomie und Astrophysik, Universit{\"a}t
T{\"u}bingen, D-72076 T{\"u}bingen, Germany}

\author{Kostas D. Kokkotas}

\affil{Department of Physics, Aristotle University of Thessaloniki,
Thessaloniki 54006, Greece}

\and

\author{Nikolaos Stergioulas}

\affil{Max Planck Institute for Gravitational Physics,
The Albert Einstein Institute,
D-14473 Potsdam, Germany}

\begin{abstract}

We present a case study for the relevance of the r-mode
instability for accreting compact stars. Our estimates are based
on approximations that facilitate back-of-the-envelope
calculations. We discuss two different cases:
 1) For recycled millisecond pulsars we argue that the
r-mode instability may be active at rotation periods longer than
the Kepler period (which provides the dynamical limit on rotation)
as long as the core temperature is larger than about $2\times10^5$
K. Our estimates suggest that the instability may have played a
role in the evolution of the fastest spinning pulsars, and that it
may be presently active in the recently discovered 2.49~ms X-ray
pulsar SAX~J1808.4-3658 as well as the rapidly spinning neutron
stars observed in low mass X-ray binaries. This provides a new
explanation for the remarkably similar rotation periods inferred
from kHz quasi-periodic oscillations in the LMXBs. The possibility
that the rotation of recycled pulsars may be
gravitational-radiation limited is interesting because the
gravitational waves from a neutron star rotating at the
instability limit may well be detectable with the new generation
of interferometric detectors. 2) We also consider white dwarfs,
and find that the r-mode instability may possibly be active in
short period white dwarfs. Our order of magnitude estimates (for a
white dwarf of $M=M_\odot$ and $R=0.01 R_\odot$ composed of
$C^{12}$) show that the instability could be operating for
rotational periods shorter than $P\approx 27-33$ s. This number is
in interesting agreement with the observed periods ($>28$ s) of
the rapidly spinning DQ Herculis stars. However, we find that the
instability grows too slowly to affect the rotation of these stars
significantly .

\end{abstract}

\section{Introduction}

The recent discovery of a new instability in rotating relativistic
stars has revived interest in problems concerning mechanisms that
may limit the rotation rate of neutron stars. As was first shown
by one of us (Andersson 1998), the so-called r-modes (Papaloizou
and Pringle 1978) are formally unstable at all rates of rotation
in a perfect fluid neutron star (see also Friedman and Morsink
1998). The mechanism behind the instability is the familiar one
that was discovered by Chandrasekhar (1970), Friedman and Schutz
(1978): The modes are unstable due to the emission of
gravitational waves  (see Stergioulas (1998) for a review on
nonaxisymmetric instabilities in rotating relativistic stars).
Even more surprising than the actual existence of the instability
are the results of recent estimates of the time-scale at which the
r-mode instability will grow (Lindblom, Owen and Morsink 1998;
Andersson, Kokkotas and Schutz 1998). It seems that the r-mode
instability has the potential to slow down newly born neutron
stars dramatically, and that the r-mode instability is
considerably stronger than the previously considered one for the
f-mode of the star (cf. the results of Lindblom 1995; Stergioulas
and Friedman 1998). The new results suggest that, assuming that a
neutron star is born spinning as fast as it possibly can (at the
Kepler frequency $\Omega_K \approx 0.67 \sqrt{\pi G \bar{\rho}}$,
i.e. with a period $P_K\approx 0.5-2$ ms), the r-mode instability
will force the star to spin down to a period of roughly $15$ ms in
the first year or so. This prediction is in good agreement with
the inferred initial period for the Crab pulsar (19 ms). The
uncertainties in the present models may also accomodate the newly
discovered 16 ms pulsar in N157B (Marshall et al 1998). The birth
spin period for this rapidly spinning pulsar is less certain, but
one can estimate that it ought to have been shorter than 9 ms
(this follows by assuming the braking index to be in the range
observed for young pulsars together with the youngest likely age
of 5000 years). Anyway, if the r-mode instability is active in a
newly born neutron star it will force the star to spin down. The
large fraction of the initial rotational energy that is then
radiated away as gravitational waves may well be detectable with,
for example, LIGO (Owen et al 1998).

These are undoubtedly exciting suggestions, that indicate that the r-mode
instability plays an important role in astrophysics. Given the
promise of the suggested scenario for newly born neutron stars,
we are inspired to investigate whether the instability
can be relevant also in other situations. In this paper
we assess the potential relevance of the instability
for accreting stars, both neutron stars and white dwarfs.
Specifically, we provide rough estimates of the
 timescales involved and ask whether the r-mode instability
provides an upper limit on the rotation of
stars spun up by accretion. We compare our estimates
to observed data for the recycled millisecond pulsars (MSPs) and
the
rapidly spinning neutron stars in low-mass X-ray binaries (LMXBs),
as well as
the DQ Herculis white dwarfs.

\section{Back of the envelope timescales}

At the present time there are no fully relativistic studies of
the dynamics of perturbed rotating stars (although the neutral modes
that signal the onset of instability of the f-mode have been
calculated, see Stergioulas and Friedman (1998)  for a fully
relativistic calculation and Yoshida and Eriguchi (1997)  for a
calculation in the relativistic Cowling approximation).
Until such calculations become available we will not know the detailed
effects of a gravitational-wave
instability. However, we
can obtain useful estimates that should be qualitatively correct,
and may not differ too much from the quantitative truth, in a
straightforward way (Lindblom et al 1998; Andersson et al 1998).
In this paper we will use the estimates obtained by
Kokkotas and Stergioulas (1998) in the context of uniform density stars
to assess the relevance of the r-mode in various
situations. We prefer to work within this approximation since the
way that the various results scale with the parameters of the star
(mass and radius) is then clear.

The uniform density approximation leads to
the gravitational radiation growth time:
\begin{equation}
t_{gw} \approx  22 \left( {1.4 M_\odot\over M} \right)
\left( {\mbox{10 km} \over R} \right)^4 \left( {P \over \mbox{1 ms}}
\right)^6 \mbox{ s} \ ,
\label{gwapp}\end{equation}
for the  $l=m=2$ r-mode
(that leads to the strongest instability).
$M$, $R$ and $P$ are the mass, the radius and the period of the star,
respectively. This estimate of $t_{gw}$ is roughly a factor of two
smaller than the results for the $N=1.0$ polytropic equation of state
and the specific stellar parameters ($M=1.5M_\odot$ and $R\approx 12.5$ km)
considered by both
Lindblom et al (1998) and Andersson et al (1998).
In a way, this discrepancy illustrates the uncertainties
associated with the present studies of the r-mode instability.
Furthermore, since models of realistic neutron stars
typically have a mean polytropic index in the range $0.5-0.8$, the difference of
a factor of two or so (between polytropes of index 0 and 1) in the gravitational
radiation timescale can be seen to
represent the uncertainty in the real equation of state of neutron star matter.
Also, as we will argue below, the difference of a factor of two has little effect on
the conclusions of the present discussion. Consequently, we will work with
the uniform density approximation (\ref{gwapp}) in the remainder of this paper.

The damping timescale due to shear viscosity in  a normal fluid star is
\begin{equation}
t_{sv} \approx 1.2 \times 10^4 \left( {1.4 M_\odot \over M} \right)^{5/4}
\left( {R \over \mbox{10 km}} \right)^{23/4} \left( {T \over 10^7 \mbox{ K}}
\right)^2  \mbox{ s}\ ,
\label{svapp}\end{equation}
where $T$ is the temperature of the star, and we have used the
shear viscosity coefficient
\begin{equation}
\eta_{nf}  = 2\times 10^{22}
\left( \rho \over 10^{15} {\rm g/cm}^3 \right)^{9/4}
\left( { T \over 10^7 {\rm K}}\right)^{-2} \mbox{ g cm}^{-1}\mbox{ s}^{-1} \ .
\label{eta}\end{equation}
This assumes that the main contribution to the shear
viscosity arises from neutron-neutron scattering, which should
be a valid approximation at temperatures above that
which the star becomes superfluid (Flowers and Itoh 1979).
In a similar way to the gravitational-wave timescale,
the normal fluid shear viscosity timescale (\ref{svapp})
has a factor of two uncertainty.

When the outer core of the star has become superfluid the main contribution to the viscous
dissipation is due to electron-electron scattering. Then
the relevant viscosity coefficient and timescale are:
 \begin{equation}
\eta_{sf}  = 6\times 10^{22}
\left( \rho \over 10^{15} {\rm g/cm}^3 \right)^2
\left( { T \over 10^7 {\rm K}}\right)^{-2} \mbox{ g cm}^{-1}\mbox{
s}^{-1} \ ,
\label{sfeta}\end{equation}
and
  \begin{equation}
t_{sv} \approx 3.6 \times 10^3 \left( {1.4 M_\odot \over M} \right)
\left( {R \over \mbox{10 km}} \right)^5 \left( {T \over 10^7 \mbox{ K}}
\right)^2 \mbox{ s}\ .
\label{sfapp}\end{equation}
For a superfluid star we should also be concerned with the so-called
mutual friction. Specifically, it has been suggested that
mutual friction will completely suppress the instability of the f-mode
(Lindblom and Mendell 1995). At present there is no similar calculation
for the r-mode instability, but it would not be surprising if mutual friction
also has a strong effect on the r-modes. However,
Mendell (1991) has suggested that mutual friction is dominated by the shear viscosity
[as given by (\ref{sfeta})] at temperatures below about $10^7$ K. This would indicate
that the effect may not be dominant in the oldest stars.
Given that results for the effect of mutual friction
on the unstable r-modes are outstanding we will not include this
effect in the present discussion. For superfluid stars we
will only account for the shear viscosity due to electron-electron scattering.
This means that our results may overestimate the strength of the instability
in the superfluid case.

On the other hand, one would not expect the entire
star to become superfluid: A cold neutron star will have an
extended solid crust. This may be highly relevant for the
r-mode estimates: The mode used in the calculation of the estimated
timescales
$t_{gw}$ and $t_{sv}$ is in many ways similar to the f-mode
of the star (albeit with a toroidal rather than spheroidal angular
dependence). Specifically, both modes are well described by
eigenfunctions
that grow as $r^l$ towards the surface (in this sense they are
``global'' modes). For this reason, we
expect them to be affected in a similar way by the presence of a
crust, and as was shown by McDermott et al (1988) the crust
has hardly any effect on the frequency of the f-mode (basically since
the mode propagates at the speed of sound which is generally much
higher than the speed of a shear wave in the crust).
We expect that this is likely to hold also for the r-mode
under consideration, and therefore the above timecales may, in fact, be
reasonable also for a star with a superfluid core and a solid crust.

As long as we are mainly
interested in relatively low temperatures, and other, more exotic,
dissipation mechanisms do not play a role,
the effect of the r-mode instability can be inferred from $t_{gw}$ and
$t_{sv}$. At high temperatures
(above $10^9$ K)
one expects the bulk viscosity of the fluid to play a
significant role, cf. the estimates of Lindblom et al (1998) and
Andersson et al (1998). However,
in accreting systems  the temperature
should always be low enough that it is sufficient to  include
only $t_{gw}$ and $t_{sv}$. When this is the case one can
easily deduce the critical rotation period above which the
r-mode instability affects the rotation of the star (when
the gravitational-wave growth time is shorter than the viscosity
damping time). We thus find
that the mode grows if the period is shorter than
\begin{equation}
P_c \approx 2.8 \left( {R \over \mbox{10 km}} \right)^{39/24}
\left( {1.4 M_\odot
\over M} \right)^{1/24} \left( {T \over 10^7 \mbox{ K}} \right)^{1/3}
\mbox{ ms} \ ,
\label{pcrit1}\end{equation}
for a normal fluid star, and
\begin{equation}
P_c \approx 2.3 \left( {R \over \mbox{10 km}} \right)^{3/2} \left(
{T \over 10^7 \mbox{ K}} \right)^{1/3} \mbox{ ms} \ ,
\label{pcrit2}\end{equation}
when we use the viscosity due to
electron-electron scattering in a superfluid. Interestingly, these
critical
 periods are not strongly dependent on the mass of the star. Furthermore,
 the uncertainties in (\ref{gwapp}),
(\ref{svapp}) and (\ref{sfapp})
have little effect on the critical period.
For example,  the uncertain factors of two in $t_{gw}$ and $t_{sv}$
individually lead to an uncertainty of 12\% in $P_c$. When combined, the
uncertainties suggest
that we may be (over)estimating the critical period at the 25 \% level.
Considering uncertainties associated with the various realistic equations
of state for supranuclear matter, and the many approximations upon which our
present understanding of the r-mode instability is based, we feel that it is
acceptable to work at this level of accuracy.

\section{Implications for millisecond pulsars}

We will now discuss the possibility that the r-mode instability
may be relevant for the period evolution of the fastest observed
pulsars. All observed millisecond pulsars have periods larger than
the 1.56 ms of PSR1937+21, and it is relevant to ask whether there
is a mechanism that prevents a neutron star from being spun up further
(eg to the Kepler limit) by accretion. Specifically, we
are interested in the possibility that the r-mode instability
plays such a role. Before proceeding with our discussion,
 we recall that Andersson et al (1998)
have already pointed out that the instability has
implications for the formation of millisecond pulsars (albeit in an
indirect way).  Specifically, the strength of the r-mode
instability seems to rule out
the scenario in which millisecond pulsars (with $P<5-10$ ms)
are formed as an immediate result
of accretion induced collapse of white dwarfs.
Continued accretion would be needed to reach the shortest
observed periods. In other words, all millisecond
pulsars with periods shorter than (say) 5-10~ms
should be recycled.

Our main question here is whether it is
realistic to expect the instability to be relevant also for
older (and in consequence much colder) neutron stars.
Even though the critical
period is much shorter for a cold star
our estimates  (\ref{pcrit1}) and
(\ref{pcrit2}) are still above the Kepler period ($\approx0.8$~ms for
our canonical star) which suggests
that the instability could be relevant.
As an attempt to answer the question we will confront our
rough approximations with observed data for MSPs, and
the neutron stars in LMXBs.

\subsection{The millisecond pulsars}

In this section we discuss the r-mode instability in the context
of the recycled MSPs. These stars are no longer
accreting, and supposing that they have been cooling for some time
they should not be affected by the instability at
present.
Our main question is whether the observed data is in
conflict with a picture in which the r-mode instability
halted accretion driven spin-up at some point in the past.

First of all, our estimates show that the rotation will be limited
by the Kepler frequency (using $P_K\approx 0.8$~ms for a canonical
star) if the interior of the star is colder than $T\approx 2\times
10^5$ K. Also, it is straightforward to show that, in order to
``rule out'' the instability (to lead to a critical period equal
to the Kepler period at {\bf e.g.} temperature $4\times10^8$ K)
the dissipation coefficient of the shear viscosity (or any other
dissipation mechanism) must be almost six orders of magnitude
stronger than (\ref{sfeta}) .

\begin{figure}[t]



\resizebox{\hsize}{10.5cm}{\includegraphics{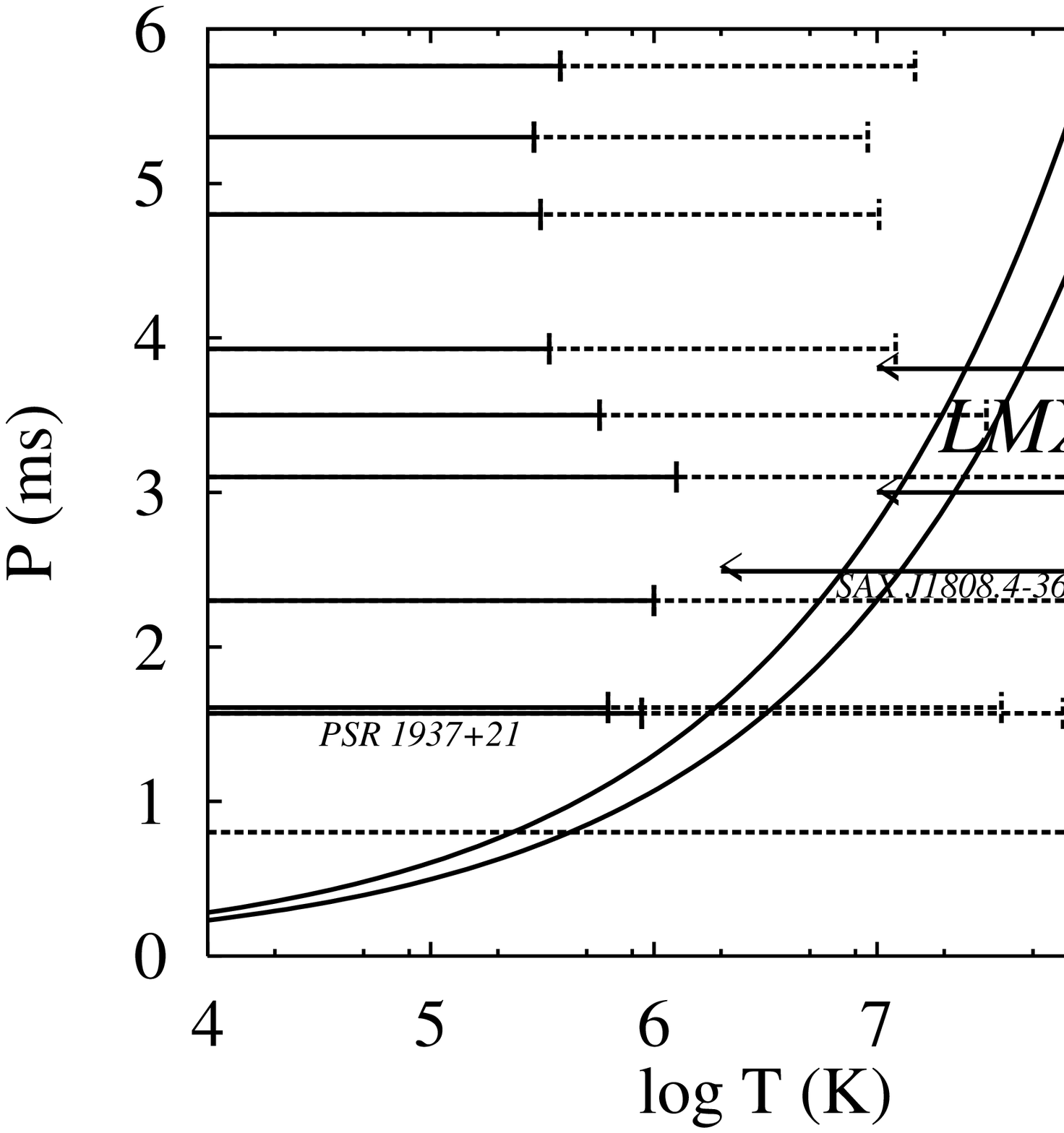}}

\caption{The inferred critical period for the
r-mode instability at
temperatures relevant to older neutron stars (solid lines).
The upper line is for a normal fluid star, while the lower one is for
a superfluid star (only taking electron-electron scattering into
account, see the main
text for discussion).
The data is for a neutron star with  $M=1.4 M_\odot$ and
$R= 10$ km. The Kepler limit, which corresponds to $P\approx 0.8$ ms
for our canonical star, is shown as a horizontal dashed
line. We compare our theoretical result to
i) the
 observed periods and temperatures of the most rapidly spinning MSPs
(see Reisenegger 1997 for the data). The surface temperatures are
indicated as solid vertical lines. The dashed continuation of each
line indicates the estimated core temperature. ii)
the observed/inferred periods and temperatures for accreting neutron
stars in LMXBs, and iii) the recently discovered 2.49~ms X-ray pulsar
SAX~J1808.4-3658.}
\label{fig1}\end{figure}

Our inferred critical periods, (\ref{pcrit1}) and (\ref{pcrit2})
(for a canonical neutron star)
are illustrated, and compared to observed periods and upper limits on
the surface temperatures (from ROSAT observations; see data
given by Reisenegger (1997)) for the fastest millisecond pulsars
in Figure~\ref{fig1}. In the figure we also indicate the associated upper limits
on the core temperatures, as estimated using eqn (8) of Gudmundsson, Pethick
and Epstein (1982).

The illustrated r-mode instability estimates would be in conflict with
the MSP observations if the interior temperature of a
certain star were such that it was placed considerably below the
critical period for the relevant temperature. Basically,
an accreting star whose spin is limited by the r-mode
instability would not be able to spin up far beyond
the critical period, since the instability would
radiate away any excess accreted angular momentum. As the
accretion phase ends, the star will both cool down and spin down
(the timescales for these two processes, photon cooling and
magnetic dipole braking, are such that a MSP would evolve
almost horisontally towards the left in Figure~\ref{fig1}).

Given the uncertainties in the available data we do not think
the possibility that the r-mode instability may have played a role
in the period evolution of the fastest MSPs can be ruled out.
First of all, it must be remembered that the
ROSAT data only provide upper limits on the
surface temperature, and the
true temperature may well be considerably lower than this.
If the true core temperatures
of the fastest spinning pulsars
were roughly one order of magnitude lower than the ones indicated
in Figure~\ref{fig1}
our inferred lower limit on the spin
period would, in fact,  be in quite good agreement with observations.
One must also recall the many simplifying assumptions
that went into our estimates of the critical period.
(Recall that we have already suggested that the uncertainties
in the critical period may be at the level of 25\%.)
The most crucial of these assumptions regards the
detailed role of superfluidity. It is not at all unlikely
that superfluid mutual friction, or some other exotic dissipation mechanism,
serves to suppress the instability and decrease the critical period.
In conclusion, Figure~\ref{fig1} is suggestive. It seems plausible that
the r-mode instability is relevant for neutron stars spun up to
millisecond periods. Furthermore, the figure
indicates how improved temperature observations should be able to test
present and
future (more detailed) models for the r-mode instability.

\subsection{Neutron stars in LMXBs}

In the standard scenario it is assumed that MSPs have evolved from
LMXBs, in which they spin up to the present rotation rate
by accreting matter from their low-mass ($<0.4M_\odot$) companions.
If the accretion rate is close to the Eddington limit
($10^{-8}M_\odot/\mbox{yr}$)
it would typically take $10^7$ years to spin a neutron star up to
a period of a few milliseconds.
Recent observations have provided interesting support
for this model. First of all, the observed kHz quasiperiodic
oscillations
(QPOs) detected by the Rossi X-ray timing explorer (see van der Klis
(1997) for a review)
show that the neutron stars in LMXBs are spinning rapidly.
Secondly, the recent discovery of the 2.49~ms X-ray pulsar
SAX~J1808.4-3658
(Wijnands and van der Klis 1998)
provides the first evidence of an evolutionary link
between the LMXBs and the MSPs.

The observations of kHz QPOs from systems containing rotating neutron
stars provide a wealth of interesting information.
Of specific relevance for our present discussion is the suggestion that
 many neutron stars in  LMXBs are spinning
with almost identical periods. This suggestion follows from the fact
 that the separation between the two KHz QPOs that are commonly observed
stays constant even though the individual
peak frequencies vary. The proposed
explanation --- the beat-frequency model (see van der
Klis (1997) for references) ---
is that this robust frequency is the underlying rotation
frequency (or a multiple thereof). Assuming that
this is the case one finds that
the observed QPOs indicate a typical rotation frequency in the range
260--330 Hz ($P$ in range 3--3.8 ms).
(Interestingly, Backer (1998) recently suggested that MSPs are typically born
with a period of 3~ms after spin-up.)
 There is, however,  one major caveat to
this interpretation of the  QPOs: For Sco X-1 the peak separation
varies considerably with the luminosity of the system. It has, in
fact, been argued that the available data is consistent with such
a variation in all LMXBs (Psaltis et al 1998). If this is the case
the peak separation cannot correspond to the rotation frequency
(since the rotation of Sco X-1 would then have to vary between
3.2~ms and 4.4~ms on short timescales), but even so it seems
reasonable that the observed beat-frequency is close (but not
equal) to the rotation frequency. Hence, we believe it makes sense
to assume that the neutron stars in the LMXBs are actually
spinning at almost identical periods and try to understand why
this is the case.

Two explanations for this
phenomenon have been proposed. In the first model (White and Zhang 1997)
the star reaches
an equilibrium period when the spin-up torque due to accretion is
balanced by the
spin-down torque due to the magnetic field. For a magnetic field dominated by
the dipole the standard equilibrium period follows from
\begin{eqnarray}
P_{eq} &=& 0.86 \left( {B\over 4\times10^8 {\rm G}} \right)^{6/7}
\left( {M \over 1.4M_\odot} \right)^{-5/7} \nonumber \\
&\times&
\left( {10^{-8} M_\odot/{\rm yr} \over \dot{M}} \right)^{3/7}
\left( {R \over 10 {\rm km}} \right)^{16/7} \mbox{ ms} \ .
\label{peq}\end{eqnarray}
Within this model one can use the
observed periods and luminosities (which imply the accretion rates)
and deduce a maximum magnetic field for each star.
This results
(cf. Table 1 of White and Zhang (1997)) in magnetic fields
very similar to those found for MSPs.
This is suggestive, but the magnetic fields deduced for the LMXBs will
only compare favourably to the observed values for the MSPs if the
field does not decay as the neutron star evolves. However, this is in
accord with the favoured model at present: The magnetic field is only
expected
to decay considerably
during the accretion phase (Romani 1990; Urpin, Geppert and Konenkov
1998).
Hence, the values that White and Zhang
(1997) infer for $B$ are reasonable as long as the low-field systems
are close to the end of the accretion phase (which could well be the
case since they are the ones with the lowest X-ray luminosity).

However, the model of
White and Zhang only leads to similar equilibrium periods if there
is some intrinsic association between $\dot{M}$
and $B^2$ [see (\ref{peq})] and, as was pointed out by Bildsten
(1998),
no such relation is seen in the data for the (slower rotating)
highly magnetized X-ray pulsars. In view of this, Bildsten proposed
an alternative explanation for the similar
rotation periods in the LMXBs.
He argued that the rotation may be limited by gravitational
radiation in these systems. Specifically, he suggested that
temperature gradients
in the star lead to a time-dependent quadrupole moment, i.e. emission
of gravitational waves that balance the spin-up torque due to
continued accretion. This is an interesting suggestion, especially
since the resultant gravitational waves may well be detectable.

It is interesting to compare the data for the LMXBs to our inferred
critical period for the r-mode instability. In order to do this, we need
to estimate the interior temperature of the neutron stars in LMXBs.
For a star that is accreting
close to the Eddington limit
hydrogen (and possibly helium) burning of the accreted material
can heat the core of the star up to a temperature
considerably higher than the typical one for non-accreting neutron stars
(Fujimoto et al 1984; Brown and Bildsten 1998).
The suggested core temperature for a rapidly accreting neutron star
is in the range $1-3\times10^8$ K. We use this estimate as an upper
limit and compare the resultant data to our critical periods in Figure~\ref{fig1}.
The Figure suggests that the r-mode instability provides a
possible explanation for the inferred
periods in the LMXBs.

Regarding the comparison between our critical periods and the LMXB
data in Figure~\ref{fig1} it is worthwhile making a few remarks.
First of all, it is not impossible that the interior temperature of the LMXB
neutron stars is actually lower than the $1-3\times10^8$~K we used as
our upper limit. As was shown by, for example, Fujimoto et al (1984)
a pion condensate in the core of the star can act as a heat sink and
keep the bulk of the star at $10^7$~K or so, even though the outer
layers of the star are heated up further by  hydrogen burning.
It is relevant to point out that
our critical period would be in perfect agreement with such a model.
Secondly, it is worth pointing out that
 the shear viscosity must be 200 times stronger than our superfluid
value (\ref{sfeta}) in order for the
r-mode instability not to be active in these
systems. If this were the case we would have
$P_c\approx 3$ ms at $T\approx 3\times10^8$ and the r-modes
would be stable in all indicated LMXBs.
Finally, it should  be noted that
the r-mode explanation is in qualitative agreement with a
possible
trend seen in the LMXB: The neutron star can only spin up further as
the accretion slows down.
In our picture (and also in Bildsten's
explanation) the star must accrete slower, and cool down,  in order
to be able to spin up. This means that
the evolution towards periods shorter than 3~ms should take place
on a cooling time-scale. In contrast, in
the model of White and Zhang (1997) the star would spin up
on the timescale that the
magnetic field decays. In reality, these two timescales
are, of course, likely to be linked in some way.

Having suggested an alternative
explanation for the clustering of the observed rotation
periods in LMXBs (that the r-mode instability prevents the
neutron stars from spinning up once they reach the critical
period) one would perhaps expect us to attempt to rule out
the other two  models. However,
given the uncertainties in all the suggested scenarios, we do not believe
it is meaningful to rule out either possibility at
the present time. To be able to do so we need more detailed modelling.
In fact,
we can see no reason why
several of these mechanisms could not be operating at the same
time, or be relevant in different individual systems.
As one can immediately deduce from (\ref{pcrit2}) and (\ref{peq})
the final outcome depends
on a delicate balance. For example, in order for the r-mode
instability
to dominate over the magnetic dipole we need $P_{eq}<P_c$.
Using (\ref{pcrit2}) and (\ref{peq}) we can translate this
into a relation between the temperature, the accretion rate
and the magnetic field strength. From this relation we find (for our
canonical star): i) For accretion at the Eddington rate
and a temperature of $10^8$~K, the r-mode instability dominates for
$B<3\times10^9$~G, i.e. for most of the observed systems. ii) For a
lower
accretion rate, 1\% of the Eddington limit, at $10^7$~K we find that
the r-modes would be dominant for $B<1.3\times10^8$~G, i.e. only a few
of the observed systems. This
simple example shows that the two mechanisms
are active at roughly the same level and that neither of them
should be ruled out at the present time.

\subsection{Gravitational-wave estimates}

As we have seen, the rotation of the fastest observed neutron stars
in accreting systems may be limited by the r-mode instability.
Hence, it is meaningful to estimate the amplitude of the gravitational
waves that could be emitted from these systems.
As was first pointed out by
Wagoner (1984), a gravitational-wave instability in an accreting star
will lead to periodic gravitational waves (assuming that
the system reaches a steady state where the excess
angular momentum of the accreted matter is radiated as gravitational
waves). Following Wagoner (1984) we can deduce
the gravitational-wave amplitude by using the angular momentum
gained by accretion:
\begin{equation}
\dot{J}= \sqrt{GMR}\dot{M}= - {m\over \omega}\dot{E} \ ,
\end{equation}
(where $\omega$ is the angular frequency of the mode) in the standard
flux formula
\begin{equation}
h^2 = {4G \over c^3} \left( {1\over \omega r} \right)^2 |\dot{E}| \ ,
\end{equation}
where $r$ is the distance to the source, and $h$ is the (dimensionless)
gravitational-wave amplitude.
Using these  relations, together with our r-mode estimates, we find
\begin{eqnarray}
h &\approx& 2.3\times10^{-26} \left( {P_c\over 1\mbox{ ms}} \right)^{1/2}
\left( {M\over 1.4M_\odot}\right)^{1/4}
\left( {R \over 10 \mbox{ km}}\right)^{1/4}  \nonumber \\
&\times& \left( {\dot{M} \over
10^{-8}M_\odot/\mbox{yr}}\right)^{1/2}
\left( {1 \mbox{ kpc} \over r} \right) \ .
\label{gwest}\end{eqnarray}

To illustrate what this means, let us provide a specific example.
Since it would provide one of the strongest sources we consider
the particular case of Sco~X-1 (cf. Bildsten 1998). Assume that
the true rotation period of Sco~X-1 is 4~ms, and that the interior
temperature is such that the r-mode instability is active and
provides a limit on the spin of the star (this corresponds to
assuming that $T\approx 10^8$~K cf. Figure~\ref{fig1}). Given
this, and the relevant data for Sco~X-1 ($r=0.7$~kpc and an
average accretion rate of $\dot{M}\approx 3\times
10^{-9}M_\odot/\mbox{yr}$ (Lang 1992)) we find that (\ref{gwest})
implies $h\approx 3.5\times10^{-26}$ for a star with canonical
mass and radius. The effective amplitude achievable after matched
filtering improves as the square-root of the number of detected
cycles, so we get $h_{eff}= \sqrt{\omega t_{obs}/2\pi} h \sim
10^{-21}$ after about two weeks of integration. After one year of
integration, a similar source at the centre of our galaxy (at 10
kpc) would reach the same effective amplitude. In other words, the
gravitational waves from galactic neutron stars whose rotation is
limited by the r-mode instability should be observable by, for
example, LIGO. The detection of this kind of essentially
continuous signals is obviously more complicated than we have
indicated here, but since the data analysis issues are identical
to those for gravitational waves from a slightly deformed rotating
neutron star we refer the interested reader to detailed studies of
that problem (Jaranowski, Krolak and Schutz 1998, Brady et al
1998).

Interestingly, given a detection of gravitational waves from the LMXBs
one would easily be able to distinguish between the
r-mode instability
and radiation from the time-dependent quadrupole model suggested by Bildsten (1998).
The r-modes radiate at $4/3$ of the rotation frequency, while a
rotating quadrupole should generate radiation of (mainly) twice the rotation
frequency. Supposing that the rotation frequency is known from a QPO
one should have no trouble distinguishing between the two mechanisms.

Finally, we should also mention that there may exist systems that
are radiating gravitational waves according to the prescribed
scenario, but are otherwise invisible.  As was suggested by Schutz
(1997) the accretion rate may  be close to the Eddington limit if
the neutron
star
is part of a so-called Thorne-Zytkow object, i.e. when it
spirals inside a giant star at the
late stages of binary evolution. Such systems should be similar
to the example we discussed above, and would likely be
detectable through the radiated gravitational waves.

\section{Implications for white dwarfs}

Although it is clear that the r-modes would formally be unstable
also in less compact perfect fluid stars,
it is not at all obvious
that the timescales will work out in such a way that the instability
has any relevance. Hence, it is interesting to attempt to extend our
estimates to white dwarfs. To achieve a rough estimate for the critical
period
is rather straightforward. We first
rescale
our expression for the growth time-scale
due to gravitational waves (\ref{gwapp})
to canonical values for a white dwarf. This leads to
\begin{equation}
t_{gw} \approx 3.1\times10^9 \left( {M_\odot \over M } \right)
\left( {0.01R_\odot \over R} \right)^4 \left( { P\over 30s} \right)^6
\mbox{ yrs} \ .
\label{wdgrav}\end{equation}
To estimate the role of the viscosity we
need to replace (\ref{eta})  with an expression that is
 relevant at white dwarf densities. As a rough approximation, we
will use (for a star composed of $C^{12}$)
\begin{equation}
\eta= 9.7\times 10^5 \left( { \rho \over 10^6 {\rm g/cm}^3} \right)^{7/4}
\left( { 10^5 {\rm K} \over T}\right)^{1/4} {\rm g cm}^{-1} \mbox{s}^{-1} \ .
\label{wdeta}\end{equation}
This expression is deduced from data used by Durisen (1973) and estimates the
electron contribution to the shear viscosity at densities
$10^4 \leq \rho ({\rm g/cm}^3)\leq 10^6$. We have used the
results for non-relativistic electrons since they lead to a larger
value of $\eta$, in order not to underestimate the role
of the viscosity. We have compared our approximation to more
recent (and detailed)
results of Itoh et al (1987). The comparison suggests that our
expression (\ref{wdeta}) is a slight overestimate of the strength
of the electron viscosity (by a factor of less than 5 in the
temperature/density range of interest).
Furthermore, we have checked that the electron contribution
dominates that of the ions in the temperature/density range of
interest. This is true as long as the density is above
$10^4$ g/cm$^3$ or so.
At lower densities the results of Itoh et al (1987) indicate that
the ion contribution will dominate over the electron contribution.
Then we expect (\ref{wdeta}) to be an underestimate by (again)
less than a factor of 5.
In conclusion,  we think that estimates obtained using (\ref{wdeta})
are reasonable, but the potential erring factors of 5 should be
kept in mind. Such errors would affect the critical rotation period at the
25\% level.  Since the main objective of the present
estimates is to assess whether a more detailed study of the role
of the r-mode instability in white dwarfs is warranted
we believe it is reasonable to work at this level of accuracy.

Using (\ref{wdeta}) we find that
\begin{equation}
t_{sv} \approx 1.7\times10^9 \left( {M_\odot \over M } \right)^{3/4}
\left( {R \over 0.01R_\odot} \right)^{17/4} \left( { T\over 10^5 {\rm
K}} \right)^{1/4}
\mbox{ yrs} \ ,
\label{wdsv}\end{equation}
and a
critical period
\begin{equation}
P_c \approx 27 \left( {M
\over M_\odot} \right)^{1/24} \left( {R \over 0.01 R_\odot}
\right)^{11/8}
 \left( {T \over 10^5 \mbox{ K}} \right)^{1/24}
\mbox{ s} \ .
\label{pwd}\end{equation}
As in the case of neutron stars,
the critical period is not strongly dependent
on the mass. Furthermore, in this case it is also weakly
dependent on the temperature. The question is: Does the
estimated critical period indicate that
the r-mode instability is relevant also
for white dwarfs?

First of all,
we should compare the deduced critical period to the
mass-shedding limit provided by the
Kepler frequency. The value of $P_c\approx 27$ s for a canonical
white dwarf (at $T=10^5$ K) should then be compared to the Kepler
period for a white dwarf: $P_K\approx 17$ s (that follows from
$\Omega_K = 0.67 \sqrt{\pi G \bar{\rho}}$,
 but if the star is differentially rotating at
the
relevant temperature the limiting period can be much smaller
than this value).
The fact that $P_c>P_K$ indicates
that the r-mode instability should become active
before a white dwarf that is spinning up reaches
the mass shedding limit.

\begin{figure}[t]



\resizebox{\hsize}{10.5cm}{\includegraphics{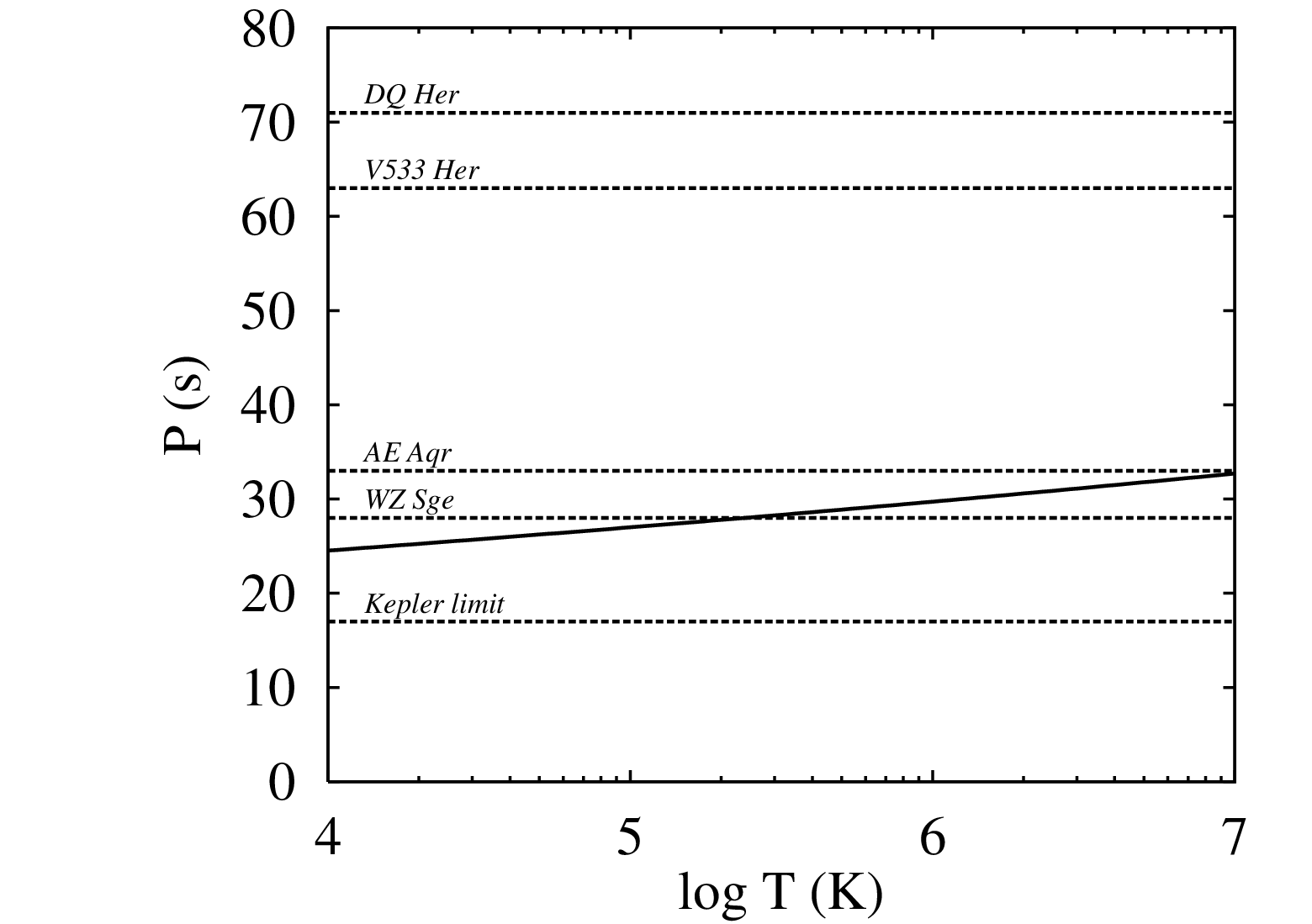}}

\caption{The inferred critical period for the
r-mode instability at
temperatures relevant for white dwarfs  (thick solid line).
The data is for a (Carbon based) white dwarf with
$M=M_\odot$ and $R=0.01R_\odot$.
The theoretical result is compared to the Kepler limit (thick dashed
line)
which corresponds to $P\approx  17$ s, and the
observed periods of the  most rapidly spinning
DQ Herculis stars (dashed lines)  (see  Patterson 1994).
In order from below, these stars are WZ Sge, AE Aqr, V533 Her and DQ Her.
The likely surface temperature of the
fastest
rotators is 15000 K , and the core temperature (which is the relevant one for the
r-mode estimates) is likely to be something like two orders of magnutide higher.
}
\label{fig2}\end{figure}

Moreover, a comparison of our result to the available
observations of rapidly spinning white dwarfs
is suggestive. There are no observations of
white dwarfs rotating faster than the limit set by (\ref{pwd}).
But the fastest spinning white dwarfs, the so called DQ Herculis stars
(which are magnetized accreting cataclysmic variables)(Patterson
1994), have periods quite close to $P_c$. The shortest
definite rotation period that has been observed is the 33 s of AE
Aqr. A likely candidate for slightly faster
rotation is WZ Sge with a period of 28 s. The estimated surface temperature
of these stars is 15000 K, and an  upper limit (above which
the colour of the stars would change considerably) is 50000 K
(Patterson, private communication). Just as in the case of neutron
stars, the interior temperature (which should be used in the r-mode estimates)
is likely to be considerably higher (maybe two orders of magnitude) than the
surface value. However, since the critical period $P_c$ is very weakly
dependent on $T$ an increase of the temperature by two orders of magnitude
only changes the critical period by 20\% (at $T=10^7$~K we get $P_c\approx 33 $ s).
In Figure~\ref{fig2} we compare our estimated critical period to the observations.

Our estimates suggest that the fastest spinning white
dwarfs that have been observed are close to the limit where
the r-modes would be unstable. This is an interesting result,
but it is clear from (\ref{wdgrav}) that the
r-modes would grow very slowly in these stars. To
see whether we should expect the instability to beable to prevent the star
from spinning up we compare the growth timescale to
the estimated spin-up time for a white dwarf accreting at the rates
observed in the DQ Her.
The spin-up timescale can be approximated by
\begin{equation}
t_{su} \approx {J \over \dot{J}} \approx
{2MR^2\Omega \over 5\sqrt{GMR}} {1\over \dot{M}} \ ,
\end{equation}
which leads to
\begin{eqnarray}
t_{su} &\approx& 1.3\times10^9 \left({ M\over M_\odot } \right)^{1/2}
\left( {R\over 0.01R_\odot} \right)^{3/2} \left( {30 {\rm s}\over P}
\right)
\nonumber \\
&\times&
\left( { 10^{-10} M_\odot/{\rm yr} \over \dot{M} } \right) \mbox{ yrs} \ .
\label{spinup}\end{eqnarray}

Interestingly, the two timescales $t_{gw}$ and $t_{su}$
would balance at $P\approx 30$ s for the observed accretion rate of AE Aqr
($\dot{M}=2\times10^{-10}M_\odot/{\rm yr}$).
Still, we believe the growth rate of the mode is
too slow for the instability to be relevant.
If the star spins up closer to the Kepler period the mode grows
faster, but at
for $P=17$ s we still get $t_{gw}\approx 10^8$  yrs.
This means that the r-mode has to be able to grow coherently
for 100 million years in order for the  instability to affect the
rotation of a white dwarf, which seems unlikely. Hence, we can probably rule
out the r-mode instability from playing any role
role in the scenario where a weak-field white dwarf is spun up towards
the Kepler frequency and then undergoes accretion induced collapse.
As was shown by Narayan and Popham (1989) accretion
induced
collapse is likely to occur only if the accretion rate is higher
than $4\times10^{-8}M_\odot/{\rm yr}$. The white dwarf is expected to
reach critical mass after accreting $0.1-0.15M_\odot$, which means
that it would collapse after $10^7$ years or so. This would clearly not
give the r-modes time to grow by the many orders of magnitude required
for the instability to
affect the rotation of the star.

Finally, a word of caution is in order. Our estimate for the viscosity
assumed that the stellar fluid was in the liquid phase. At low
temperatures this may not be the case, and the viscosity of the
resultant
lattice  may be much higher than (\ref{eta}). We estimate
that the instability  becomes active
 unless the true viscosity of white dwarfs
is more than 400 times larger than
(\ref{wdeta}).

\section{Final remarks}

We have presented rough estimates that suggest that
the r-mode instability may be active in accreting
neutron stars and white dwarfs.
Our back-of-the-envelope calculations lead to
two main conclusions.

Firstly, the r-mode instability
can limit the rotation of recycled MSPs provided that
they are hotter than $2\times10^5$ K.
We have shown that this suggestion cannot be ruled out by present
observations for MSP periods and temperatures.
We have also compared our predicted critical periods to observed
data for LMXBs. In particular, we considered
the systems for which a rotation period
in the narrow range 3--3.8~ms has been inferred from the
beat-frequency
model for QPOs. This comparison strongly suggests that the r-mode instability
may be active in these systems, and that it can potentially be
the agent that limits the rate of rotation to the observed periods.
Furthermore, we have estimated the amplitude of the
gravitational waves from
a neutron star rotating at the r-mode instability limit
and accreting at the Eddington limit (typical examples could be the
fastest accreting LMXBs or  the so-called Thorne-Zytkow objects).
The result suggests
that this kind of source (from within our Galaxy) could be detected by
the new generation of gravitational-wave detectors.

There are, of course, many uncertainties associated with
these suggestions. The most important one concerns the basically unknown
role of superfluidity. A better understanding of neutron star
superfluidity and its effect on the r-mode instability is
urgently needed to firm up the present estimates.

Secondly, we find that
the r-mode instability may be active in
short period white dwarfs. Our order of magnitude
estimates (for a $C^{12}$ white dwarf of $M=M_\odot$ and $R=0.01
R_\odot$) yield
a critical period $P\approx 27-33$ s.
This number is in interesting agreement with the observed
periods ($>28$ s) of the rapidly spinning DQ Herculis stars.
It is, however, only an order of magnitude estimate that indicates
that the instability should be taken seriously also for white dwarfs
and that a more detailed study should be attempted.

\section*{Acknowledgements}

We thank J. Friedman, J. Patterson and B.F. Schutz for helpful
discussions. We would also like to thank the anonymous referee
whose comments helped improve the paper considerably. We are
grateful to the Max-Planck-Institute for Gravitational Physics
(Albert-Einstein-Institute) in Potsdam, for generous hospitality
during the workshop ``Neutron star dynamics and gravitational-wave
emission'' (May 11--15 1998) and to all the participants for
discussions.

\pagebreak

\begin{references}


\reference{r1} Andersson  N. 1998  {\em Ap. J} {\bf 502} 708
\reference{r1b} Andersson N., Kokkotas K.D. and Schutz B.F. 1998
{\em Gravitational radiation limit on the spin of young neutron
stars} to appear in {\em Ap. J.} [astro-ph/9805225]
\reference{bac} Backer D.C. 1998 {\em Ap. J.} {\bf 493} 873
\reference{bild} Bildsten L. 1998 {\em Ap. J. Lett.} {\bf 501} 89
\reference{brady} Brady P., Creighton T., Cutler C. and Schutz
B.F. 1998 {\em Phys. Rev. D} {\bf 57} 1
 \reference{bb} Brown E.F. and Bildsten L. 1998 {\em Ap. J.}
{\bf 496} 915 \reference{r2} Chandrasekhar S. 1970 {\em Phys. Rev.
Lett.} {\bf 24} 611 \reference{che} Cheng K.S., Chau W.Y., Zhang
J.L. and Chau H.F. 1992 {\em Ap. J.} {\bf 396} 135 \reference{du}
Durisen R.H. 1973 {\em Ap. J.} {\bf 183} 205 \reference{fi}
Flowers E. and Itoh N. 1979 {\em Ap. J.} {\bf 230} 847
\reference{r4} Friedman J.L. and Morsink S. 1998  {\em Ap. J.}
{\bf 502} 714 \reference{r5} Friedman J.L. and Schutz B.F. 1978
{\em Ap. J.} {\bf 222} 281 \reference{fuj} Fujimoto M.Y., Hanawa
T., Iben Jr I. and Richardson M.B  1984 {\em Ap. J.} {\bf 278} 813
\reference{gud} Gudmundsson E.H., Pethick C.J. and Epstein R.I.
1982 {\em Ap. J. Lett.} {\bf 259} 19 \reference{ikt} Itoh N.,
Kohyama Y. and Takeuchi H. 1987 {\em Ap. J.} {\bf 317} 733
\reference{jara} Jaranowski P., Krolak A. and  Schutz B.F. 1998
{\em Phys. Rev. D} {\bf 58} 063001 \reference{ks} Kokkotas K D and
Stergioulas N 1998 {\em Analytic description of the r-mode
instability in uniform density stars}, {\em A\&A}, in press,
[astro-ph/9805297]

\reference{lang} Lang K.R. 1992 {\em Astrophysical Data} (Springer
Verlag, New York)

 \reference{r9} Lindblom L. {\em Ap. J.} {\bf 438}
265 (1995) \reference{r10} Lindblom L. and Mendell G.  1995 {\em
Ap. J.} {\bf 444} 804 \reference{r11} Lindblom L., Owen B.J. and
Morsink S.M. 1998 {\em Phys. Rev. Lett.}  {\bf 80} 4843
\reference{mm} Marshall F.E., Gotthelf E.V., Zhang W., Middleditch
J. and Wang Q.D. 1998  {\em Ap. J. Lett.} {\bf 499} 179
\reference{mcd} McDermott P.N., Van Horn H.M and Hansen C.J. 1988
{\em Ap. J.} {\bf 325} 725 \reference{mend} Mendell G 1991 {\em
Ap. J.} {\bf 380} 530 \reference{np} Narayan R. and Popham R. 1989
{\em Ap. J.} {\bf 346} L25 \reference{ow} Owen B, Lindblom L.,
Cutler C., Schutz B.F., Vecchio A. and Andersson N. 1998 {\em
Phys. Rev. D} {\bf 58} \reference{r13} Papaloizou J. and  Pringle
J.E. 1978 {\em MNRAS} {\bf 182} 423 \reference{pa} Patterson J.
1994 {\em Publ. Astron. Soc. Pac.} {\bf 106} 209 \reference{r14}
Provost J., Berthomieu G. and Rocca A. 1981 {\em Astron. Astrop.}
{\bf 94} 126 \reference{psa} Psaltis D., M{\'e}ndez M., Wijnands
R., Homan J., Jonker P.G., van der Klis M., Lamb F.K., Kuulkers E,
van Paradijs J. and Lewin W.H.G. 1998 {\em Ap. J. Lett.} {\bf 501}
L95 \reference{reis} Reisenegger A 1997 {\em Ap. J.} {\bf 485} 313
\reference{rom} Romani R.W. 1990 {\em Nature} {\bf 347} 741
\reference{sc} Schutz B.F. 1997 pp 11--17  in {\em Mathematics and
Gravitation. II. Gravitational wave detection} ed: A. Krolak
(Banach Center publications vol 41, Warzaw) \reference{r16}
Stergioulas N. and Friedman J.L. 1998 {\em Ap. J.} {\bf 492} 301
\reference{S98} Stergioulas N. 1998 {\em Rotating Stars in
Relativity}, {\em Living Reviews in Relativity}, Vol. 1, 1998-8,
http://www.livingreviews.org \reference{urp} Urpin V., Geppert U.
and Konenkov D. 1998 {\em MNRAS}, {\bf 295} 907 \reference{yosh}
Yoshida S. and Eriguchi Y. 1997  {\em Ap. J.} {\bf 490} 779
\reference{vdk} van der Klis M. 1996 in {\em The many faces of
Neutron stars} \reference{wa} Wagoner R.V. 1984 {\em Ap. J.} {\bf
278} 345 \reference{wz} White N.E. and Zhang W. 1997 {\em Ap. J.}
{\bf 490} L87 \reference{wv} Wijnands R. and van der Klis M. 1998
{\em The discovery of the first accretion-powered millisecond
X-ray pulsar} [astro-ph/9804216]
\end{references}
\end{document}